# Decay of high-energy electron bound states in crystals

Tadas Paulauskas[1,2], Robert F. Klie [1]

[1]Department of Physics, University of Illinois at Chicago, Chicago, IL, USA

[2]Center for Physical Sciences and Technology, Vilnius, Lithuania

## Abstract

High-energy electrons that are used as a probe of specimens in transmission electron microscopy exhibit a complex and rich behavior due to multiple scattering. Among other things, understanding the dynamical effects is needed for a quantitative analysis of atomic-resolution images and spectroscopic data. In this study, state-correlation functions are computed within the multislice approach that allow to elucidate behaviors of transversely bound states in crystals. These states play an important role as a large fraction of current density can be coupled into them via focused electron probes. We show that bound states are generically unstable and decay monoexponentially with crystal depth. Their attenuation is accompanied by a resonant intensity transfer to Bessel-like wavefunctions that appear as Laue rings in the far-field diffraction patterns. Behaviors of bound states are also quantified when thermal effects are included, as well as point defects. This approach helps to bridge the Bloch wave and multisliced electron propagation pictures of dynamical scattering providing new insights into fundamental solutions of the wave equation, and may assist in developing quantitative STEM/TEM imaging techniques.

## Introduction

The relativistically-corrected Schrodinger equation, as can be derived from the Dirac equation, is widely used to model high-energy electron propagation in crystals. Elastic scattering is arguably one of the fundamental mechanisms by which structural information is retrieved in (scanning) transmission electron microscopy, (S)TEM. Dynamical scattering also affects the spatial localization of electronic excitations as captured in the electron energy-loss spectroscopy. Numerical simulations are required in many cases to assist image interpretation and to perform quantitative analysis. Furthermore, ensuing advancements in instrumentation, including electron detectors and high-energy electron optics, require that theory and computational technique developments also go in tandem [1-6].

Extensive theoretical and numerical work on elastic and thermal scattering of high-energy electrons in crystals can be found in the field of transmission electron microscopy. General approaches include the S-matrix, Bloch wave, and the multislice [7-10]. There are also variations and hybrid methods tailored for inelastic scattering simulations and electron propagation in non-periodic systems[11-13]. Solutions of the wave equation reflect crystal symmetries resulting in Bloch waves. It is known that such high-energy electron Bloch waves, in particular their wavefunctions transverse to the high-momentum direction, can be spread out or delocalized, and also dispersionless core-like states bound to atomic-columns. Boundary conditions at the crystal surface determine which states are populated. Theoretical work has shown that in STEM a highly-focused electron beam can couple a large fraction of the current density into $s$-states bound to individual atomic-columns. Subsequent dynamics in conjunction with thermal effects has been suggested to be responsible for the observed contrast formation in high-angle annular dark-field (HAADF) and annular bright-field (ABF) imaging modes [14-18].

The $s$-state models discussed in literature typically neglect 3-dimensional treatment of the bound states. Their excitation coefficients are assumed to be constant, but due to quasi-elastic thermal scattering phenomenological exponential decay factors are usually added. In this study, we apply new computational methods within the multislice approach to extract the high-energy transversely bound states and compute state-correlation functions. This allows to elucidate behaviors of bound states within a purely elastic scattering regime and their link to formation of Laue rings. Thermal effects in the frozen-phonon approximation and scattering from point defects in a form of substitutional atoms are also examined within this framework [19-21].

This article is organized as follows. We review the numerical multislice approach to be used in the computational section. In order to analyze results presented in the computational part, we propose a model theoretical framework describing the high-energy electron atomic-column states. New computational techniques are then implemented within the multislice code and are used to quantify behaviors of bound states. Practical applications of the findings are discussed.

## 1. The wave equation and the multislice method

We seek solutions to the wave equation describing a high-energy spinless electron in an external electrostatic crystal potential. The usual ansatz is taken by factoring a large momentum plane wave along the optic-axis and a slower varying component, $\Psi(r,z) = \psi(r,z)e^{2\pi i k_z z}$ [7,10]. Wavefunction $\psi(r,z)$ then satisfies the following Schrodinger-like equation:

$$\frac{i}{\sigma}\frac{\partial}{\partial z}\psi(r,z) = \left[-\frac{\hbar^2}{2\gamma m_0}\nabla_r^2 - V(r,z)\right]\psi(r,z) \tag{1.1}$$

Here, $r=(x,y)$ is a Cartesian vector transverse to the optic-axis $z$ which represents electron's primary direction of motion. The interaction constant: $\sigma = \gamma m_0/k_z \hbar^2$ with units (energy· distance)$^{-1}$; relativistic factor $\gamma = 1 + |e|V_0/m_0 c^2$, where $|e|, V_0$ is the electron's charge and the accelerating voltage, respectively. This is the starting point in the traditional multislice approach. The subsequent numerical implementation can be arrived at in several ways, such as the original Cowley and Moodie physical optics treatment, formal quantum mechanical operator method as described in [10], among others [21,22]. Here, we follow a formal integral approach [23]. One notes that terms in the brackets of Eq. (1.1) represent a transverse Hamiltonian with $z$-dependent potential. Eq. (1.1) has a mathematical form of a time-dependent Schrodinger equation and a general integral solution can be written as:

$$\psi(r,z) = \mathcal{T} e^{-i\sigma \int_0^z ds\, H(r,s)} \psi(r,0) \tag{1.2}$$

Above, $\mathcal{T}$ is the $z$-ordering operator for the transverse Hamiltonian $H(r,z)$ in the brackets of Eq. (1.1). The traditional multislice algorithm results from "Troterrization" (Suzuki-Trotter approximation) of the $z$-evolution operator in Eq. (1.2). In particular, $z$-axis is discretized with small steps Δz, possibly of different sizes, and also some sort of an approximation to the continuous electrostatic potetial $V(r,z)$ is applied. A straightforward approach is to integrate an atomic potential along the z-axis and replace it with a constant value for an extent Δz. This translates to a subdivision of an entire crystal into slices stacked normal to the large momentum direction. The $z$-variation over each slice is replaced by a mean potential by averaging over the slice thickness Δz:

$$V_j(r) = \frac{1}{\Delta z} \sum_i \int_{-\infty}^{\infty} dz\, V_{ij}(r - r_i, z). \quad (1.3)$$

The total electrostatic potential of a slice $j$ is a sum $i$ of atomic potentials assigned to $j$ which are individually averaged along $z$. Integration limits in Eq. (1.3) represent this particular choice of approximation. Troterrization of Eq. (1.2) thus results in ordered unitary operators $U_j$ that represent electrons' $z$-evolution over each individual slice along its path. Assuming slices of equal thicknesses, the wavefunction at a depth $N\Delta z$ in a crystal is obtained:

$$\psi(r, N\Delta z) = \prod_{j=N}^{1} U_j(r, \Delta z)\psi(r, 0) \quad (1.4)$$

$$U_j(r, \Delta z) = e^{-i\sigma H_j(r)\Delta z} \quad (1.5)$$

The transverse $z$-independent Hamiltonian operator of a $j^{th}$ slice $H_j=K+V_j(r)$, and corresponds to terms in the parentheses of (1.1), but with a potential $V_j$ in Eq. 1.3; $\Pi$ is the ordering operator for $U_j$'s. Each individual $U_j$ to first-order approximation is: $U_j(r, \Delta z) = e^{-i\sigma V_j(r)\Delta z} e^{i\sigma K\Delta z}$, or a symmetrized version is sometimes quoted as well.

We note that in Eq. (1.2) a formal solution assumes the Dyson-type series. Since transverse Hamiltonians $H_j$ of different slices $j$ in the Trotterized version (Eq. 1.4,1.5) generally do not commute, one would not expect to find stationary states induced by $z$-evolution operators $U_j$. This is not a result of the discretization, however, since it is well known that time-domain Floquet/Bloch states, also known as driven quantum systems, obeying a wave equation analogous to Eq. (1.1), exhibit resonance phenomena and decay of bound states [24-28]. This is in contrast to the Bloch wave methods which explicitly seek stationary and linearly independent states of the high-energy electrons in crystals. In the following sections, we will analyze behaviors of states induced by $z$-evolution operators in the multislice approach.

## 2. Theoretical framework

In this section, a theoretical model of atomic-column electron states is introduced. It will serve to analyze numerical results presented in Sections 3 and 4. This framework also helps to lay the ground for the numerical techniques presented in the Appendices. We note that this model framework is only approximate, since the excitation coefficients are assumed to be constant and states are linearly independent. Nevertheless, the high-energy electron bound states exhibit stationary-like behavior due to long lifetime of their resonance character, as will be shown numerically.

### 2.1 Atomic-column electron states

Starting from the paraxial Eq. (1.1), the total Hamiltonian is expressed as:

$$\mathcal{H} \equiv H(r,z) - \frac{i}{\sigma}\frac{\partial}{\partial z} \tag{2.1}$$

$$\mathcal{H}\,\psi(r,z) = 0 \tag{2.2}$$

We seek solutions $\psi$ with a short-ranged isolated atomic-column potential satisfying periodicity $V(r, z + a_z) = V(r, z)$. According to the Bloch/Floquet theorem, there are solutions $\psi$ to Eq. (2.2) of the form:

$$\psi_\kappa(r,z) = \phi_\kappa(r,z)\, e^{-i\varepsilon_\kappa \sigma z} \tag{2.3}$$

$$\mathcal{H}\phi_\kappa(r,z) = \varepsilon_\kappa \phi_\kappa(r,z) \tag{2.4}$$

The $z$-dependent functions $\phi_\kappa$ are eigenstates of the Hamiltonian $\mathcal{H}$ (Eq. 2.1) with eigenvalues $\varepsilon_\kappa$, and obey periodicity $\phi_\kappa(r, z + a_z) = \phi_\kappa(r, z)$. Without loss of generality, $z$-periodic eigenfunctions $\phi_\kappa(r,z)$ can be expanded in Fourier series. An arbitrary high-energy electron wavefunction is then decomposed into a linear superposition of (2.3) and has the following form:

$$\psi(r,z) = \sum_\kappa d_\kappa \phi_\kappa(r,z) e^{-i\varepsilon_\kappa \sigma z} = \sum_\kappa \sum_{n=-\infty}^{\infty} d_\kappa c_n^\kappa \chi_n^\kappa(r)\, e^{-ing_z z}\, e^{-i\varepsilon_\kappa \sigma z} \tag{2.5}$$

The summation with respect to $\kappa$ in (2.5) is to be understood as a sum over discrete bound states ($\varepsilon_\kappa < 0$) and an integral over the continuum of scattering states. Here, $d_\kappa$ is a complex-valued expansion factor, also called an excitation coefficient, and is for now assumed to be constant.

46Section 2.3 discusses determination of $d_\kappa$ from boundary conditions. $\chi_n^\kappa(r)$ is the $n^{th}$ transverse harmonic (or a mode) of the $\kappa^{th}$ state $\psi_\kappa$ arising from the Fourier expansion of the eigenstates $\phi_\kappa(r,z)$. The reciprocal lattice vector above $g_z = \frac{2\pi}{a_z}$. Each mode $\chi_n^\kappa$ is weighted by a complex amplitude-factor $c_n^\kappa$ such that $\sum_n |c_n^\kappa|^2 = 1$. Also note that eigenvalues $\varepsilon_\kappa$ are defined uniquely modulo $n g_z \sigma^{-1}$ for any integer $n$.

Using Dirac notation, the inner-product and orthogonality of eigenfunctions $\phi_\kappa(r,z)$ are defined as:

$$\langle\langle \phi_\kappa | \phi_{\kappa'} \rangle\rangle = \sum_{n,n'=-\infty}^{\infty} c_n^{*\kappa} c_{n'}^{\kappa'} \langle \chi_n^\kappa | \chi_{n'}^{\kappa'} \rangle \langle n | n' \rangle = \delta_{\kappa,\kappa'} \quad (2.6)$$

$$\langle n | n' \rangle = \frac{1}{a_z} \int_0^{a_z} dz \, \langle n | z \rangle \langle z | n' \rangle = \frac{1}{a_z} \int_0^{a_z} dz \, e^{i n g_z z} e^{-i n' g_z z} = \delta_{n,n'} \quad (2.7)$$

$$\langle \chi_n^\kappa | \chi_n^{\kappa'} \rangle = \int_{-\infty}^{\infty} dr \, \langle \chi_n^\kappa | r \rangle \langle r | \chi_n^{\kappa'} \rangle = \int_{-\infty}^{\infty} dr \, \chi_n^{*\kappa}(r) \chi_n^{\kappa'}(r) = \delta_{\kappa,\kappa'} \quad (2.8)$$

Although here we remain in the configuration space $\mathbb{R}^3$, the above form of eigenfunctions $\phi_\kappa$ and the inner-product can be seen to represent a composite state-space $Z \otimes \mathcal{R} = \mathbb{L}^2[0, a_z] \otimes \mathbb{L}^2(\mathbb{R}^2)$. A double bra-ket notation in Eq. (2.6) is used to distinguish an inner-product, or a matrix element, in $Z \otimes \mathcal{R}$, as opposed to a single bra-ket used for Z (Eq. 2.7) or $\mathcal{R}$ (Eq. 2.8). Note that modes $|\chi_n^\kappa\rangle$ labeled by different $n$ and either same or different $\kappa$ are generally not linearly-independent in transverse $\mathcal{R}$-space, i.e. $\langle \chi_{n'}^\kappa | \chi_n^\kappa \rangle \neq 0$.

## 2.2 General remarks

To build a picture of the dynamics described above, it is worthwhile first to consider a Hamiltonian which does not depend on z, $\widehat{H}(r)$, e.g. averaged continous column-string potential is used. The wavefunction expansion (Eq. 2.5) becomes $\psi(r,z) = \sum_\kappa d_\kappa \widehat{\phi_\kappa}(r) e^{-i\widehat{\varepsilon_\kappa}\sigma z}$. Here, $\widehat{H}\widehat{\phi_\kappa} = \widehat{\varepsilon_\kappa}\widehat{\phi_\kappa}$ and so eigenvalues $\widehat{\varepsilon_\kappa}$ are exactly the transverse electron energies corresponding to eigenfunctions $\widehat{\phi_\kappa}$. For a fixed relativistic γ-factor, the eigenfunctions $\widehat{\phi_\kappa}$ form a complete and orthonormal basis for a Hilbert space $\mathcal{M} = \mathbb{L}^2(\mathbb{R}^2)$ where solutions admit a discrete bound ($\widehat{\varepsilon_\kappa} < 0$) as well as absolutely continous scattering spectrum ($\widehat{\varepsilon_\kappa} > 0$), for an isolated column. This is the starting point for traditional *1s*-state models, which can be recovered in the current framework if coefficients $c_{n\neq 0}^\kappa = 0$ in the expansion Eq. (2.5).

The electrostatic potential is felt weakly along z at large energies and so the periodic column-potential can be considered as a small perturbation to the fully-projected counter-part above. We then expect that $\chi_0^\kappa(r) \approx \widehat{\phi_\kappa}(r)$ and $\varepsilon_\kappa \approx \widetilde{\varepsilon_0^\kappa} = \langle\langle\chi_0^\kappa|H(r,z)|\chi_0^\kappa\rangle\rangle$, which is a condition that $c_{n\neq 0}^\kappa \ll c_0^\kappa \ \forall n$. Eventhough this condition is met for many bound states, as we will find numerically, nevertheless, *n=0* modes $\chi_0^\kappa(r)$ are not eigenfunctions of $H(r,z)$ (see Eq. 2.1,2.2). The equivalent of the transverse energy $\widehat{\varepsilon_\kappa}$ is now instead the expectation value of $H(r,z)$ in an eigenstate:

$$\langle\langle\phi_\kappa|H|\phi_\kappa\rangle\rangle = \varepsilon_\kappa + \sum_{n\in\mathbb{Z}}|c_n^\kappa|^2 \frac{n\,g_z}{\sigma} \quad (2.9)$$

Thus, $\varepsilon_\kappa$ represents the transverse energy averaged over one period $a_z$ (see eqs. 2.6-2.8). The summation term drops-out for symmetrically distributed modes. Based on the expansion in Eq. (2.5), we can assign to each mode $\chi_n^\kappa$ a characteristic wavevector $k_{\kappa,n} = -\varepsilon_\kappa\sigma - ng_z$, where the quantization in discrete $ng_z$ represents a crystal momentum. An eigenstate $\phi_\kappa$ can thus be viewed as a coherent superposition of states $|\chi_n^\kappa\rangle$ which carry probabilities $|c_n^\kappa|^2$ to have exchanged $n$ momentum-quanta with the crystal, as encoded in $|n\rangle$, and the spatial form of such a 'kicked' state represented by $\chi_n^\kappa(r)$. Each eigenstate has the same total energy *E,* which can be approximately separated here into the longitudinal and transverse parts: $E = m_{\perp,\kappa}c^2 + \varepsilon_\kappa$, where the so-called transverse mass: $m_{\perp,\kappa}c^2 = ((\hbar c k_\kappa)^2 + (m_0 c^2)^2)^{1/2}$ [29]. The states can thus be also qualified behaving as scattering(bound) if $E - m_{\perp,\kappa}c^2 > 0$ (<0).

### 2.3  Excitation of electron states in crystals

At the crystal surface *z=0*, the incident probe $\psi_i$ is represented by a 2-dimensional transverse wavefunction. We decompose an initial state into a superposition of crystal electron states, Eq. (2.5), but with longitudinal components all in the 'ground' state before an interaction with a crystal occurred, i.e. $c_{n\neq 0}^\kappa = 0$: $|\psi_i\rangle\rangle = \sum_\kappa d_\kappa|\chi_0^\kappa\rangle|0\rangle$, where $d_\kappa = \langle\psi_i|\chi_0^\kappa\rangle$. This implies that an electron probe $\psi_i$ excites states $\psi_\kappa$ in a crystal by coupling only via their *n=0* modes (zeroth order Laue zone). Our numerical simulations supports the claim that by using an exact $\chi_0^\kappa(r)$ mode as the input, only this single atomic-column bound state can be excited. However, such matching of a boundary wavefunction leads to the following effect. If we take an initial state as an exact normalized mode $n=0$, the subsequent unitary dynamics implies that there should be an interval $\bar{z}$ during which a transition $|d_\kappa|^2 \to \sum_{n=-\infty}^{\infty} |c_n^\kappa|^2$ to all the harmonics $\chi_n^\kappa$ composing $\phi_\kappa$ occurs. This suggests that states in Eq. (2.5) represent steady- or asymptotic-states, which emerge in the wavefront propagation picture. As it is not clear how to analytically represent the transitioning interval (here we find $< 2$ *nm*), this effect is shown in numerical simulations Sect. 4.

## 2.4 Spatial form of wavefunctions

A subset of eigenstates that are studied here correspond to transversely bound states with $\varepsilon_\kappa < 0$. The azimuthal-invariance of an atomic column-potential implies that each transverse harmonic inherits this symmetry. It follows that $\chi_n^\kappa(r)$ amy be treated as eigenfunctions of a Hamiltonian with an *effective* 2D atomic-potential *V(r)*. The general form of such wavefunctions in polar coordinates is:

$$\chi_n^\kappa(r,\theta) = R_n^{\eta l}(r) e^{i l \theta} \qquad (2.10)$$

The radial wavefunctions $R_n^{\eta l}$ are characterized by the principal $\eta$ and the angular momentum $l$ quantum numbers, $\kappa \to \eta l$. The angular-part $e^{i l \theta}$ is an eigenfunction of the orbital angular-momentum operator $L_z$, which can be labeled by $l=0,1,2,3...$ as *s,p,d,f*....[30]. States with $l \neq 0$ are doubly degenerate, although this degeneracy may be lifted within a crystal. Atomic-column bound states are thus called *1s*, *2s*, *2p*, etc., and should not be confused with orbitals of atoms.

Given the localized nature of bound states, weakly-bound states in a full crystal potential could be constructed following the tight-binding formalism [31]. Eigenvalues $\varepsilon_{\kappa,k_\perp}$ will then generally differ from the isolated-column states:

$$\varepsilon_{\kappa,k_\perp} = \varepsilon_\kappa + \sum_{r_m \neq 0} e^{i k_\perp \cdot r_m} \alpha(r_m) \qquad (2.11)$$

The term $\varepsilon_\kappa$ corresponds to the eigenvalue of an isolated-column state, but which now also includes a correction from the full-crystal potential (e.g. inner-potential). The second term adds the transverse $k_\perp$-dependence, weighted with overlap integrals $\alpha(r_m)$.

## 3. Electron states in SrTiO₃ [100] at 200 $keV$

In this section, eigenvalues and eigenfunctions of high-energy electrons are computed within the multislice propagation scheme in SrTiO₃ (STO) along <100> crystallographic axis at primary 200 $keV$ energy. Doyle and Turner scattering factor parametrizations are employed, and with 0.01 Å pixel size for multislice propagations [32,33]. The numerical techniques implemented within the multislice, auxiliary spectral filtering and quantum diffusion methods, are described in more detail in Appendix 1. Excitation coefficients and state-correlation functions are analyzed in Sect. 4, and in Appendices. We start by examining *isolated* Sr and Ti-O atomic-column states of <100> STO structure, and then present results from simulations in a full STO crystal. Bound state behaviors are also explored in the frozen-phonon model, whereby STO columns are taken as a collection of atoms statically-displaced from their equilibrium positions.

### 3.1 Subslicing

The STO crystal structure viewed down <100> is composed of a motif of three distinct atomic columns, pure Sr and O columns as well as a mixed Ti-O column. The unit cell along the beam direction is naturally sliced twice separating atoms within 1.9525 Å slabs. Note that higher harmonics, $\chi_n^\kappa(r)e^{-ing_z z}$, (Eq. 2.5) have associated longitudinal frequencies such that to sample them, and faithfully extract wavefunctions $\chi_n^\kappa$, the propagation steps in the multislice algorithm need to be finer. There is no unique way to accomplish this, and in fact, a subslicing scheme pertains to the accuracy of a scattering approximation from individual atoms, since steps become a small fraction of an atomic potential's range [21]. The method adopted in this article simply takes the two distinct periodic STO slices and subdivides propagation over their thickness into four identical substeps. In particular, propagation over each individual *z*-constant slice (Eq. 1.4,1.5) reduces to $U_j(\Delta z) = \left[ U_j\left(\frac{\Delta z}{m}\right) \right]^m$, and $U_j\left(\frac{\Delta z}{m}\right) = e^{-i\sigma V_j(r)\Delta z/m}e^{i\sigma K\Delta z/m}$, where *m=4* and $\Delta z$=1.9525 Å. The choice of *m* in regards to the numerical implementation is discussed further in the Appendix 1. Another way to perform the subslicing is to use a full crystal potential and sub-integrate given depth ranges into discrete thin slices. Both of these methods were tested, showing no fundamental differences in the observed spectral weights and bound state dynamics. However, scattering intensity from individual atoms (e.g. to HAADF range) with different subslicing/subpropagation schemes slightly differs. The eigenvalues $\varepsilon_\kappa$ were also observed here to converge with an increasing number of subslicing steps.

### 3.2 Mode spectrum of Sr column *1s*-state

Figure 1 a) shows the longitudinal wave number spectrum of a single high-energy electron Sr *1s* column-state. The spectrum is obtained by computing a finite Fourier transform with respect to depth *z* of the state-correlation function and plotting its absolute value (Appendix). We use the before-hand extracted Sr *1s* *n=0* mode $\chi_0^{1s}(r)$ as an input state to propagate within the multislice code so as not to excite any other Sr column bound states. The peak widths are determined by the propagation depth chosen in the calculation, as well as an additional Lorentzian broadening due to an exponential decay of the excitation coefficient (Sect. 4). The peak-heights are proportional to the *effective* weights $\widetilde{c_n} = |c_n^{1s} \langle \chi_0^{1s}|\chi_n^{1s}\rangle|$, where $c_n^{1s}$ is the weight coefficient of a given mode (see Eq. 2.5 and discussion Sect. 2.3). Positions on the horizontal wave number axis follow: $k_{\kappa,n} = \varepsilon_\kappa \sigma - n g_z$. The zero of the horizontal axis marks the incident electrons' primary wave number $k_z$, which is factored-out in the wavefunction's ansatz (Sect, 1). Hence, positive/negative values indicate anti-parallel/parallel contributions of each mode to the free-space plane wave, respectively. The spectrum gives the eigenvalue $\varepsilon_{1s} \approx -113$ *eV* of the Sr *1s*-state (*n=0* mode), or equivalently its wavenumber $\varepsilon_{1s}\sigma$=-0.08 Å$^{-1}$. The effective weight ratios for the two strongest modes are $\frac{|\widetilde{c_1}|}{|\widetilde{c_0}|} \approx 0.06$ and $\frac{|\widetilde{c_3}|}{|\widetilde{c_0}|} \approx 0.01$. The actual weight coefficients $|c_n^{1s}|$ are somewhat larger due to the overlap integral contribution, for example, we find $\langle \chi_0^{1s}|\chi_{n=1}^{1s}\rangle \approx 0.8$. The peaks of *+/-n* modes were found to be symmetric (in terms of peak height) regardless of an input and reference states used in the calculations. The relative peak heights are also influenced by *z* Fourier coefficients of the crystal potential.

### 3.3 Sr *1s*-state with thermal displacements

Figure 1b) shows the spectrum of Sr *1s*-state obtained by including a single configuration of transverse thermal atomic displacements within the frozen-phonon model. Thermal displacements obey Gaussian distribution with isotropic 0.05Å RMS radius from the equilibrium position. The spectrum suggests that despite increased distortion in the column the *1s*-state is nevertheless established during an electron's propagation. The bound states' eigenvalue is pushed towards less negative values (see also Section 4.3). Broadening in the spectra suggests that discrete Fourier expansion over the modes in Eq. (2.5) may have to be replaced by a weighted integral if small translational lattice-symmetry breaking is present. The associated *1s*-state excitation coefficient of this atomic arrangement with thermal and point defect effects is further investigated in Sect. 4.3.

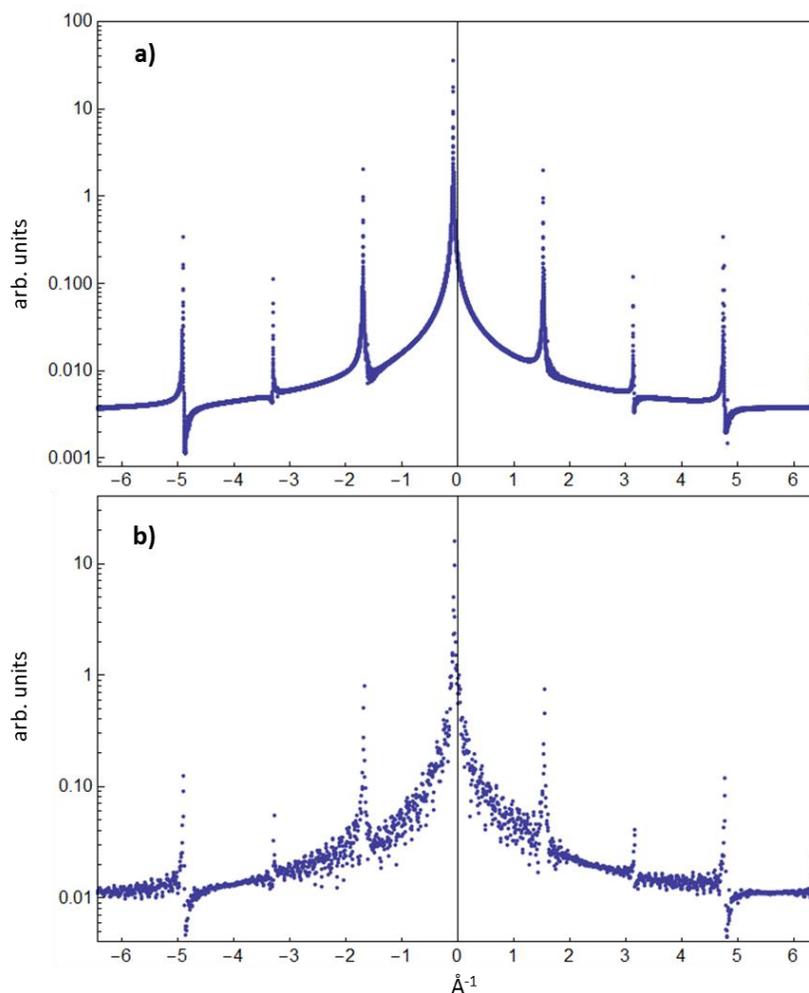

**Figure 1. Longitudinal wavenumber $k_z$ spectrum (inverse angstroms) of an individual Sr *1s*-state modes. The peak heights are directly related to the weights of the higher-order modes. a) Sr *1s* spectrum, b) Sr *1s* with one configuration of thermal displacements within the frozen-phonon model, includes three substitutional Au atoms.**

## 3.4 Bound state spectrum in [100] SrTiO$_3$ crystal

The eigenvalue spectrum $\varepsilon_\kappa$ (refers to *n=0* modes only) of a large periodic [100] SrTiO$_3$ supercell is shown in Figure 2. The spectrum was obtained by using an input wavefunction that is a sum of several Gaussians and first-order vortex beams positioned over Sr, Ti-O and O columns. Such wavefunction excites most of the distinct high-energy electron bound states in STO and the peak heights reflect this particular form of input. For example, a vortex beam carrying *l*-orbital angular momentum at the probe forming aperture position has the form $\Psi(k_\perp) = A(k_\perp)e^{il\theta}$ and can match the transverse *2p*-states (see Eq. 2.10) with suitably chosen $A(k_\perp)$, providing a large excitation coefficient. The peak assignments in Fig. 2 were given by analyzing/propagating each state individually. The labeling for weakly-bound O *1s,* Sr/Ti-O *2p, 2s* suggests only the main character of these states as they hybridize in STO to some extent with adjacent overlapping column-states. The presence of a hybridization can be tested by noting e.g. interference effects in the auto-correlation functions that use input states $\chi_0^\kappa(r)$ of the corresponding isolated-column states and is reflected in their eigenvalue shifts (not shown here). A reason for the emergence of the peak above zero in some structures is not immediately obvious althought it is related to the nearly-free electron propagation. Copies of spectral peaks (higher-modes) in Fig. 2 are also found at positions $\varepsilon_\kappa \sigma -/+ ng_z$ (not shown here), similar to the individual column *1s*-state spectra of Fig. 1.

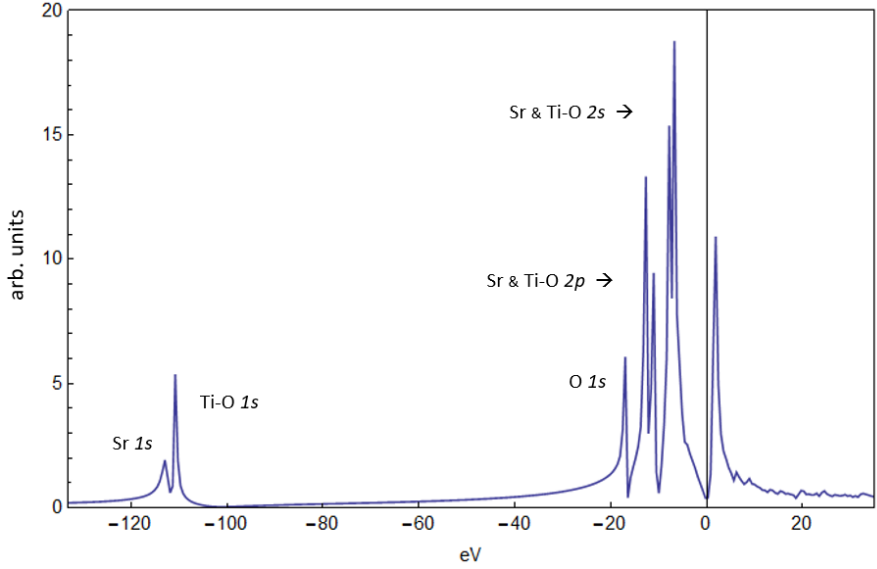

**Figure 2.** Eigenvalue $\varepsilon_\kappa$ spectrum in [100] SrTiO$_3$ crystal, which also corresponds to *n=0* modes' longitudinal wavenumbers $\varepsilon_\kappa\sigma$. The labeling of weakly bound states in the energy regime > -20 *eV* is only suggestive as per their dominant character, since theses states are hybridized to some extent within the crystal. Heavy columns Sr and Ti-O each hold three well-pronounced bound states, while pure oxygen columns support only a single *1s*-state.

### 3.4 Wavefunctions

Figure 3 shows a representative selection of transverse Sr isolated-column wavefunctions $\chi_n^\kappa(r)$. The wavefunctions are obtained with both (where applicable) quantum diffusion (method of imaginary time), and the spectral filtering (see Appendices). The results were cross-checked for consistency. Figures 3 (a),(b),(c),(g),(h) correspond to Sr $\chi_0^{1s}, \chi_{-1}^{1s}, \chi_0^{2s}, \chi_0^{2p}$ and $\chi_{-1}^{2p}$ wavefunctions, respectively. Their forms can be seen to obey Eq. (2.10). Only the *1s*-state is core-like, while higher-energy states *2s* and *2p* are more delocalized and hybridize within the STO environment, as noted in the previous section. The higher-order *n<0* mode wavefunctions are obtained with spectral filtering by 'dialing-in' their wave numbers from the spectrum calculated as in Fig. 1a). Note how much more localized are *n=-1* harmonics of *1s* and *2p* states compared to *n=0*, and the higher modes become increasingly more localized with *n*. Knowledge of relative phase factors in $c_n^\kappa = |c_n^\kappa|e^{i\theta_n}$ allows, in principle, to reassemble an entire bound state in the expansion (Eq. 2.5). The relative phase factors can be found by fitting of the state-correlation functions and also by a visual inspection of the extracted states, as in Fig. 3. For example, the Sr $\chi_{-1}^{2p}$ modes lobes are rotated by $\theta_1 = \pi/2$ compared to $\chi_0^{2p}$. This phase difference between *n=0* and *n=1* modes holds for the $\chi_n^{1s}$ states as well, shown by fitting of the state-correlation functions in the Appendix.

Wavefunctions $\chi_1^{1s}, \chi_2^{1s}, \chi_1^{2p}$ in figures 3(b),(e),(i), respectively, are obtained via spectral filtering by choosing the positive *n* harmonics' $k_z$ values. The longitudinal wave numbers of these harmonics are degenerate with states in the continuum, and the obtained wavefunctions clearly represent scattering states. The extracted isolated column wavefunctions very closely resemble so called non-diffracting Bessel beams. It can readily be checked that the transverse wavefunctions approximately fit the analytic form: $\chi_n^{nl}(r) \approx J_0(2\pi|k_{\perp,n}^\eta|r)e^{il\theta}$. Here, $J_0$ is the 0th order Bessel function of the first-kind with orbital-angular momentum $l = 1$ in (i) and 0 in Fig. 3(b),(e). We note that the analytic expression of Bessel wavefunctions $\chi_n^{nl}(r)$ is a form of a plane wave. The spatial extent of the wavefunctions obtained with the filtering technique (Fig. 3(b), (e)), more generally for finite electron propagation, are clearly finite showing slow intensity spreading into the outer rings. The wavefunctions in Fig. 3 (b),(d),(e) are plotted in Fig. 4(h) as an incoherent sum of their Fourier transformed absolute value squares: $|\chi_0^{1s}(k_\perp)|^2 + |\chi_1^{1s}(k_\perp)|^2 + |\chi_2^{1s}(k_\perp)|^2$. This shows the transverse momentum probability distribution in these states and also corresponds to their far-field diffraction patterns. In the far-field, the Bessel-like functions have the form and radii of the first and second-order rings, referred to as higher-order Laue zone features (HOLZ) [34]. The transverse wavevectors, or the radii $k_\perp$, can be read out from the plots in h): 4.42 Å$^{-1}$, 6.41Å$^{-1}$, and 4.51 Å$^{-1}$ for Sr *1s n=1, n=2* and Sr *2p* states, respectively.

In the kinematic scattering approximation, the rings are customarily explained as the Bragg diffraction due to the constructive interference of plane waves on a conical surface, and also marks position of the Ewald sphere intersecting Laue zones. In this study, we show numerically (Sect. 4) that formation of HOLZ rings also represents a dynamic transfer of intensity from excited bound states into the Bessel states, suggesting a resonance phenomena. That is, for each excited energetically distinct bound state there emerges a set of fingerprint-like HOLZ rings.

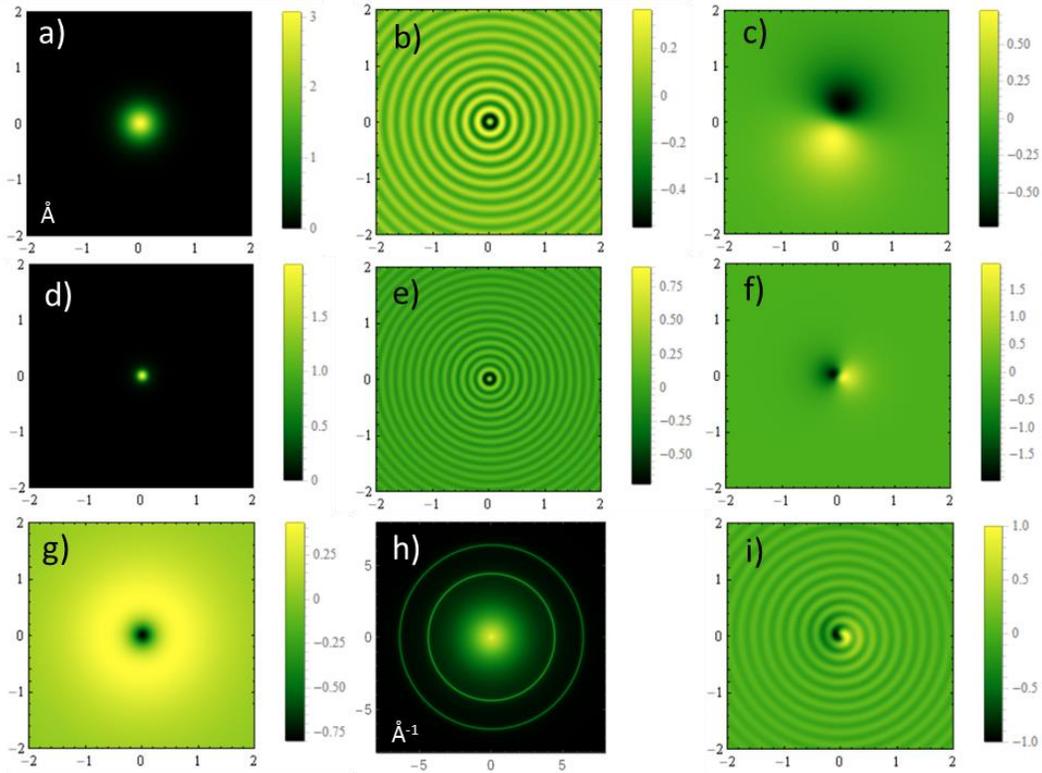

**Figure 3. a) Sr *1s n=0* wavefunctions (scale in Å), b) real-part of extracted Sr *1s* positive *n=1* mode representing a scattering state, c) real-part of Sr *2p n=0* mode, d) Sr *1s n=-1* mode, e) real-part of Sr *1s n=2* mode, representing a scattering state, f) real part of Sr *2p n=-2* mode, g) Sr *2s n=0* mode, h) sum of absolute values squared of Fourier transforms of b)+d)+e); Fig. b) gives the 1$^{st}$ HOLZ ring at $k_\perp$ =4.42 Å$^{-1}$, while e) the ring at 6.41 Å$^{-1}$, i) Sr *2p n=1* mode representing scattering state (HOLZ ring not shown, but $k_\perp$= 4.51 Å$^{-1}$).**

## 4. Dynamical evolution of bound states

By employing transverse modes of bound state eigenfunctions, such as the ones presented above, state-correlation functions are computed in this section. These complex functions relate to state excitation coefficients, while their absolute value squared is the survival probability (here also called state population), and can be tracked as a function of specimen depth. It is shown that an attenuation of a bound state is accompanied by an emergence of HOLZ rings. Individual atomic columns are examined, including effects of thermal displacements and point defects.

### 4.1 Attenuation of excitation coefficients

We use a normalized *1s*-state' *n=0* mode, $\psi(0) = \chi_0^{1s}$, extracted numerically before-hand to propagate with the multislice code and state(auto)-correlation functions are computed. Sr and Ti-O column' *1s*-state populations are then plotted: $|\mathfrak{C}_{1s}(z)|^2 = |\langle \chi_0^{1s} | \psi(z) \rangle|^2$, (Eq. 2.5). For more details refer to Appendices Eqs. (A1.1), (A2.1), and Fig. A2. In the multislice algorithm (Eq. 1.4,1.5) this formally reads as:

$$|\mathfrak{C}_{1s}(z)|^2 = | \int dr\, \chi_0^{*\,1s}(r) \prod_j e^{-i\sigma H_j(r)\Delta z} \chi_0^{1s}(r) |^2 \qquad (4.1)$$

Plots in Fig. 4 a) indicate that excitation coefficients are not constant but rather decay monoexponentially. In particular, *all modes* $\chi_n^\kappa$ composing a particular state are found to decay with the same rate so that an entire bound state decays as a whole. We also confirm from simulations that decay constants $\Gamma_\kappa$ are directly proportional to amplitudes $|c_n^\kappa|$ of the higher-harmonics. For example, the dominant mode amplitude $|c_1^{1s}|$ is ~3x larger for Sr than Ti-O *1s* state, and this shows in Fig. 4 a). Higher-order amplitudes $|c_n^\kappa|$ of a pure oxygen column *1s*-state in STO are negligibly small (not shown here) and the attenuation is accordingly negligible over typical TEM sample thicknesses. Attenuation of Sr and Ti-O isolated-columns' *2s* and *2p*-states (not shown here) is likewise much weaker than of deeply bound *1s*-states.

The non-exponential transitioning interval $\bar{z} \approx 10$ Å, more pronounced for Sr *1s*-state, can be seen in Fig. 4b), which is a zoomed-in interval from 0 to 70Å of Fig. 4a). It reflects settling-in of the higher-order modes upon the state excitation with a pure *n=0* mode, as mentioned in Sect. 2.3. This effect is more pronounced for states that have larger amplitudes $|c_{n\neq 0}^{\kappa}|$.

The *1s*-state survival probability functions plotted in Figures 4 a) and b) can be approximated as (Appendix 1, 3):

$$|\mathfrak{C}_{1s}(z)|^2 = e^{-2\sigma \Gamma_{1s} z} \left( \sum_n a_n^{1s} Cos[ng_z z + \theta_n] \right)^2 \tag{4.2}$$

Real-valued factors $a_n^{1s}$ represent effective weights, which depend on the actual weights $c_n^{\kappa}$ as well as transverse wavefunctions' overlap integrals. The sinusoidal modulations result from the higher-order modes composing a bound state vibrating into the reference state' $\chi_0^{1s}$ envelope, since $\chi_0^{1s}$ and $\chi_{n\neq 0}^{1s}$ modes are generally not orthogonal in $\mathcal{R}$.

The exponential decay term in Eq. (4.2) must arise from projections of a bound state onto a subset of the continuum of scattering states $\kappa$. In particular, terms of the form $\langle \chi_{n'}^{\kappa} | \chi_n^{1s} \rangle$ with $n' \neq n$ are expected to cause non-vanishing contributions. Provided that the overlap density approximates the Lorentzian distribution (Breit-Wigner form), the exponential decay representing a dephasing process will be recovered. Decay in the Breit-Wigner form is captured via complex eigenvalues $\varepsilon_{\kappa} \rightarrow \varepsilon_{\kappa} - i\Gamma_{\kappa}$, where decay width $\Gamma_{\kappa}$ represents a pole in the complex-plane. The independent-state description in Section 2 is hence somewhat limited, although $\Gamma_{\kappa}$ are typically small and high-energy electron bound states still retain their character, but as long-lived resonances rather than purely stationary states.

### 4.2 Intensity in HOLZ rings

The appearance of HOLZ rings in dynamical scattering is associated with an excitation of bound states [7,34]. To show quantitatively how these features are a result of bound state attenuation, we plot the intensity transfer from decaying Sr and Ti-O *1s* states into the simultaneously emerging HOLZ rings in Figure 4c). This graph is obtained by separately propagating initial states $\chi_0^{1s}$ within the multislice scheme and integrating the electron intensity $|\psi(k_\perp; z)|^2$ in the momentum-space as a function of *z* reaching thin annulus of width 1.0 Å$^{-1}$ centered on the 1$^{st}$ order HOLZ rings (Fig. 3h). Fig. 4c) accompanies the state decay plot in Fig. 4a) up to the vertical dashed line marking 200 Å thickness. Figure 4c) shows that at least 90% of the intensity decay from the Sr and Ti-O *1s* states observed in Fig. 4a) is being transferred into this narrow region around each of the 1$^{st}$ order rings. The analogous characteristic intensity transferring into other higher-order rings occurs as well but with a much smaller magnitude. The decay rate of core-like states, as far as we tested, does not change within a full crystal environment. Main difference appears to be that dephasing via the scattering Bloch waves causes modulations of the circular diffraction HOLZ ring pattern due to dispersion.

For completeness, a fundamental dynamical Bloch wave eigenvalue equation in crystals is presented in the Appendix. It is noted there that the matrix equation is non-Hermitian, although its full form is rarely used in actual calculations in the electron microscopy.

### 4.3 Bound states with thermal effects and point defects

Figure 4 d) shows the normalized intensity $|\mathfrak{C}_{1s}(z)|^2$ in Sr *1s*-state, similar to Fig. 4a), but with thermal displacements (frozen-phonon model) and also includes point defects in a form of 3 substitutional Au atoms in the column. This figure/atomic-column structure supplements the spectra plotted in Fig. 1b). The attenuation occurs much faster with thermal effects, compared to the Sr *1s* in Fig. 4a), and is less regular, although it still displays the higher-mode modulations and a clear monoexponential character. The positions of three Au substitutional atoms can be clearly seen at 50, 100 and 200 Å. The intensity beneath the reference state $\chi_0^{1s}$ drops upon encountering an Au atom (magnified inset in the figure), in proportion to the electron intensity reaching that depth. The amplitude ringing observed in $|\mathfrak{C}_{1s}(z)|^2$ after each such drop is similar in appearance to the initial state propagation/excitation at the surface *z*=0.

Table I compares thermal effects on the Sr *1s*-state eigenvalues by: 1) applying Debye-Waller factors to atomic scattering factors, 2) the frozen-phonon model thermal displacements (300*K*), and 3) without any thermal effects [35]. The values were obtained from the state spectra and fitting of the state population decays. Frozen-phonon and Debye-Waller methods show the real component of $\varepsilon$ being pushed to higher values because of the effective interaction potential broadening experienced in both cases by a propagating high-energy electron. As is expected, the Debye-Waller approach predicts a much lower state population decay which is captured in the complex component $\Gamma$. Note that the multiplicative interaction constant $\sigma$ is slightly larger at lower accelerating voltages but Table I lists only $\Gamma$. The overall decay is faster at 200 *keV* than at 1200 *keV* within the frozen-phonon model. Inclusion of thermal displacements give a smaller effect on the less deeply bound Sr and Ti-O column *2s, 2p*-states, and oxygen column *1s*-state.

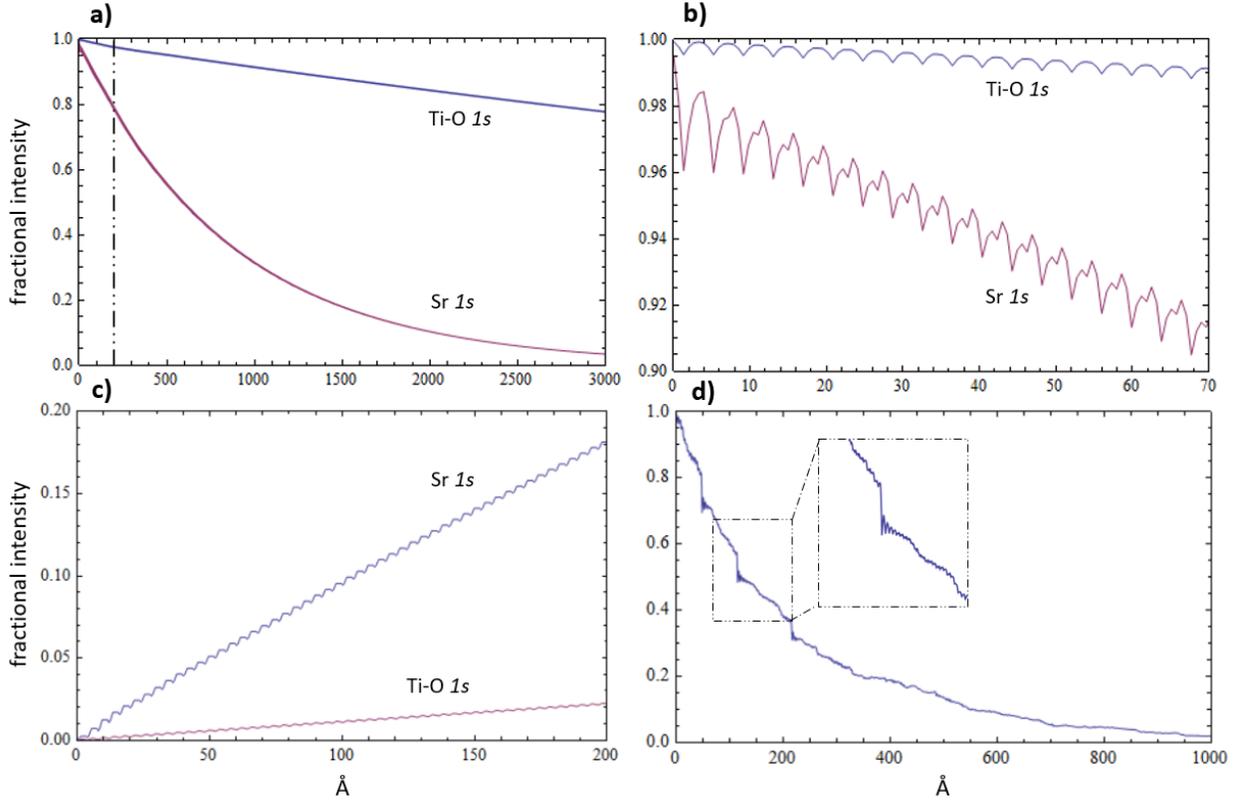

**Figure 4. a) State population (or survival probability) as a function of crystal depth, corresponds to calculation of correlation-function $|\langle \chi_0^{1s}|\psi(z)\rangle|^2$ with input and reference states both as normalized Sr and Ti-O *1s*-state modes $\chi_0^{1s}$, as indicated in the Figure. Dashed vertical line marks 200 Å, for reference to Fig. 4 c). b) Zoomed-in region of a). c) Integrated intensity (normalized) around the resonance position in momentum space in a thin annulus of width 1 Å$^{-1}$ centered around the 1$^{st}$ order Laue ring corresponding to Fig. 3h), for each *1s*-state separately. d) Thermal Sr *1s*-state population function with three Au impurity atoms, corresponds to spectrum in Fig. 1 b).**

| $E_0$ | 200 keV | | 1200 keV | |
|---|---|---|---|---|
| $\varepsilon'$ (eV) | $\varepsilon$ | $\Gamma$ | $\varepsilon$ | $\Gamma$ |
| **Debye-Waller** | -101.1 | 0.1 | -157.3 | 0.0001 |
| **Frozen-phonon** | -102.4 | 3.2 | -163.0 | 0.7 |
| **None** | -112.8 | 0.7 | -186.5 | 0.005 |

Table 1. Complex eigenvalues $\varepsilon' = \varepsilon - i\Gamma$ of Sr *1s*-state in STO [100] at 200 and 1200 *keV* energies. Obtained using Debye-Waller thermal potential broadening, frozen-phonon model atomic displacements and without any thermal effects.

## 5. Discussion

Certain conditions must be established in order for bound states to play a role in STEM/TEM imaging and EELS. Non-negligible probe current density needs to be coupled to these spatially localized states by maximizing their excitation coefficients (Section 2.3), i.e. positioning the electron probe atop a particular atomic-column, and maximizing the probes' transverse wavefunction' similarity (overlap integral) to that of bound states' at the crystal surface. It shows that present-day spherical aberration-corrected STEM instruments can couple up to ~50-60% of the intensity to *1s*-states using a suitably chosen probe convergence angle. Degenerate *2p*-states with +/- 1 $L_z$ angular momenta are coupled with approximately equal weights by an off-column positioning in STEM, or can be excited individually via focused electron vortex beams.

Findings in this work may prove useful in development of scattering potential inversion techniques, feature recognition algorithms in pixelated electron detectors, and atomic-resolution EELS quantification. For example, accuracy of a scattering potential inversion based on Bloch wave approaches declines for thick samples (>10-20 *nm*) and with strong dynamical scattering. Current work based on the robust multislice approach reveals initial dynamics at the sample surface, i.e. the complex settling of localized Bloch waves in the first few nanometers, and also helps to quantify strong HOLZ scattering components without use of phenomenological absorptive factors.

Higher-order modes' transverse wavefunctions and the central regions of resonant Bessel waves have spatial localization over atomic-columns comparable in the extent to core atomic orbitals. Hence, these dubious modes may have a non-negligible contribution to the ionization and core-loss EELS. Quantifying 3-dimensional distributions of individual dopants in bulk crystals is currently challenging for a number of reasons, including a rather coarse depth resolution achievable in STEM, and dynamical scattering effects. We briefly touched on the effect of point defects, quantifying how *1s*-state population abruptly drops at the impurity positions in proportion to state population reaching that depth. Further work is needed to investigate how this intensity redistributes, which may allow to fingerprint such impurity scattering. Pixelated electron detectors could be used to tap diffraction intensity fingerprints of individual bound states excited in a sample, as well as decay contributions towards HOLZ features, which provide crystallographic information along specimens' depth-dimension.

## Summary and Conclusions

One of the key findings in this study is a generic decay of bound high-energy electron states in crystals. The higher-order modes, analogous to the photon field-dressed states in driven quantum systems, appear to be responsible for the transitioning into the continuum of scattering states. The description of a probe-electron propagating in a crystal as a linearly independent sum of Bloch states is hence somewhat limited. However, bound states still retain their stationary-like character, but as long-lived resonances. The decay was shown here to result in a coherent intensity transfer into scattering states which form HOLZ rings. Use of absorptive scattering-factors and complex eigenvalues is commonly encountered in the phenomelogical Bloch wave, and sometimes multislice approaches, to treat thermal scattering. We found that within the frozen-phonon model bound states are still established, however, the monoexponential decay constants are enhanced. Further systematic studies will help to determine relationship between the phenomenological absorptive factors and presently observed attenuation.

The numerical techniques that were employed here are auxiliary to the multislice approach. In particular, they merely extract the transverse wavefunctions and longitudinal frequencies that are naturally established within a multislice simulation. No fine unit-cell subslicing, as was done here, is necessary, to observe the decay and HOLZ features. Nevertheless, the subslicing allows to analyse effects of higher-order modes, since these states have frequencies that require finer sampling. This article mainly focused on *1s*-states, which play a major role in the atomic-resolution HAADF and ABF characterization with modern STEM instruments, since large fraction of the probe current density can be coupled into them. However, other bound states, e.g. *2p, 2s,* molecular-like and hybridized states, are amenable to same techniques. Their contribution may prove important for other modes of imaging and spectroscopy. The framework of analysis and findings presented in this article will assist in quantitative information retrieval about a specimen via feature detection in pixelated electron detectors, efficient utilization of vortex beams and to gauge dynamical scattering effects in atomic-resolution electron energy-loss spectroscopy.

**Acknowledgements:**

The authors thank S. Findlay and J. Etheridge for the helpful discussions.

The authors acknowledge funding from the U.S. Department of Energy (EERE-DE0007545).

# Appendices

## A1. Quantum diffusion and the spectral method

The following sections describe some details how numerical methods were employed in the present study. The spectral method is used to obtain the eigenvalue spectra, as well as eigenfunctions via spectral filtering [36]. The quantum diffusion, also known as the method of imaginary time, is another technique employed to extract eigenfunctions [17]. Both these techniques can be implemented for the analysis of periodic systems within the multislice algorithm with a minimal addition to the code.

### A1.1 Spectral method – eigenvalue spectra and wavefunction filtering

An initial wavefunction that is propagated within the multislice algorithm is chosen fiducially to excite states of interest. An initial state on the sample surface can be expressed as: $|\psi(0)\rangle = \sum_\kappa d_\kappa |\chi_0^\kappa\rangle |0\rangle$. All higher modes composing a given eigenstate $\phi_\kappa(z)$ settle-in as well with fixed amplitudes $c_n^\kappa$. To obtain the wavevector $k_z$ spectrum, or the quasi-energy spectrum $k_z/\sigma$ of the form $\varepsilon_\kappa + \frac{n g_z}{\sigma}$, we first numerically calculate the state-(auto)correlation function. It is obtained by composing a list of inner-products in $\mathcal{R}$ from each slice. The state(auto)-correlation function $\mathfrak{C}(z)$ taken with respect to the initial state $\chi_0^\kappa$ assumes the following form:

$$\mathfrak{C}(z) = \langle\psi(0)|\psi(z)\rangle = \sum_\kappa \sum_{n=-\infty}^{\infty} d_\kappa^* c_n^\kappa \langle\chi_0^\kappa|\chi_n^\kappa\rangle e^{-i\varepsilon_\kappa \sigma z} e^{i n g_z z} \qquad (A1.1)$$

After some finite propagation depth $T$ the auto-correlation function is Fourier transformed with respect to $z$ giving its conjugate complex function $S(k_z)$, and has the following analytical form:

$$S(k_z) = \sum_\kappa \sum_{n=-\infty}^{\infty} d_\kappa^* c_n^\kappa e^{i\theta_n} \langle\chi_0^\kappa|\chi_n^\kappa\rangle \frac{i\,[1 - Exp[-i\,T(\varepsilon_\kappa \sigma - n g_z - k_z)]]}{(\varepsilon_\kappa \sigma - n g_z - k_z)} \qquad (A1.2)$$

By taking the absolute value (or Re, Im parts etc.) of $S(k_z)$ the bound state wavevectors $k_z$ are then laid out in a form of spectral peaks. The peak height of a given mode is proportional to the product of the complex amplitude $c_n^\kappa$ and overlap $\langle\chi_0^\kappa|\chi_n^\kappa\rangle$ and the overall excitation coefficient $d_\kappa^*$. The ket reference-state appearing in the right-hand side inner products Eq. (A1.2) can be chosen independently of the initial probe state used to excite a state(s). This allows, for example, to recover relative weights $|c_n^\kappa|$ by choosing appropriate initial trial and reference wavefunctions. Furthermore, the phase factors in $e^{i\theta_n}$ (we write $c_n^\kappa = |c_n^\kappa|e^{i\theta_n}$) can be extracted from curve fitting of the correlation function (see Appendix 2). The spectral peaks are broadened due to finite propagation depth $T$, and there is an intrinsic resolution and maximum value of $k_z$: $\Delta k_z = 2\pi/Z$, and $k_{zmax} = \pi/\Delta z$, respectively. Complex eigenvalues could also be employed to take into account an additional Lorentzian broadening due to monoexponential decay of states. To uncover modes in the spectra with non-zero $n$ requires reduction of $\Delta z$, as can be seen by comparing $ng_z$ with the expression for $k_{zmax}$. In this study, the subslicing is done according to the composition property of $U_j$'s such that $U_j(\Delta z) = [U_j(\Delta z/m)]^m$. Alternatively, a subintegration of atomic potentials can be used, which may provide a more accurate approximation to scattering from individual atoms.

The transverse wavefunction $\chi_n^\kappa(r)$ of a particular harmonic $n$ can be extracted by multiplying both sides of Eq. (2.5) with a complex-conjugate phase-factor corresponding to a particular modes' longitudinal part $e^{i(\varepsilon_\kappa\sigma + ng_z)z}$, and integrating the expression with respect to $z$ up to some finite value $T$. The analytic form of this integral provides an envelope in the $k_z$ domain identical to the one in $S(k_z)$ positioned around a chosen mode. Therefore, this method works best for well in the longitudinal domain separated states, but it also allows to extract a weighted sum of scattering states under an $S(k_z)$ envelope. The filtering is implemented numerically within the multislice approach Eq. (1.4) as:

$$|\chi_n^\kappa\rangle = \sum_j \left( \mathcal{T}_{j=1}^N U_j(\Delta z)|\psi(0)\rangle \, e^{i(\varepsilon_\kappa\sigma + ng_z)\Delta z_j} \right) \quad (A1.3)$$

The integration is replaced by summation of the propagated wavefunction, with the additional phase-factor $e^{i(\varepsilon_\kappa\sigma + ng_z)\Delta z_j}$, from all $N$ slices and subslices. The slice-ordering operator of $U_j$ is $\mathcal{T}_{j=1}^N$, and the final wavefunction still needs to be renormalized. Harmonics $\chi_{n>0}^\kappa$ of bound states tend to have longitudinal wavevectors that are degenerate to scattering states $\psi_m$ at $k_m = -\sigma\varepsilon_\kappa + n\,g_z$. Thus, filtering at these values targets resonance states.

The FFT algorithm used in the multislice imposes periodic-boundary conditions, hence the transverse numerical grid needs to be large enough to avoid wavefunction wrap-around and spurious interference effects for large propagation $T$. An absorbing smooth function with a form of a circular 2-dimensional Fermi-Dirac distribution was used here to prevent the wavefunction from reaching grid boundaries, in addition to clipping maximum transverse wavevectors to reduce aliasing errors.

## A1.2 Quantum diffusion

The quantum diffusion (QD) method, also called the method of imaginary time, when applied to high-energy electron wave equation amounts to a Wick rotation $z \rightarrow -iz$ of the optic-axis. The paraxial equation then assumes a mathematical form of the classical diffusion equation, but with density as a complex-valued function. Before analyzing application of QD to periodic states, first consider $z$-independent crystal. An electron wavefunction inside such a crystal is represented as: $|\psi(z)\rangle = \sum_\kappa d_\kappa |\phi_\kappa\rangle e^{-i\varepsilon_\kappa \sigma z}$. After the Wick-rotation ($e^{-i\varepsilon_\kappa \sigma z} \rightarrow e^{\varepsilon_\kappa \sigma z}$) and long z-propagation only the deepest bound state $\phi_\kappa$ survives. Other states $\kappa'$ are damped with decay lengths $\alpha = 1/\sigma(\varepsilon_{\kappa'} - \varepsilon_\kappa)$ relative to the ground state $\kappa$. A suitable trial wavefunction is chosen to extract a particular state $\phi_\kappa$ such that its $d_\kappa \neq 0$. In order to extract the next higher transverse-energy state the Gram-Schmidt orthogonalization procedure is employed. In particular, the orthogonal components of $m$ lower-energy states are removed from a (Wick-rotated) propagated trial wavefunctions $\psi(z)$ at each $z$ step, with inner-products taken in $\mathcal{R}$. The convergence is tracked by plotting the transverse energy expectation value, $<E> = \langle \phi_\kappa | H | \phi_\kappa \rangle$, as a function of the propagation depth. Energy variance $\Delta E$ is also a useful quantity to gauge the convergence.

The QD is applicable to $\chi_0^\kappa$ transverse wavefunctions. However, now each targeted mode is 'dressed' with all the higher-harmonics $\chi_n^\kappa$ composing the eigenstate and are being simultaneously amplified with fixed ratios. Therefore, care needs to be taken to minimize the contribution from these modes. To illustrate this effect, we plot in Figure A1 the state-correlation function propagated using the Wick-rotated axis with the multislice evolution operators $U_j$.

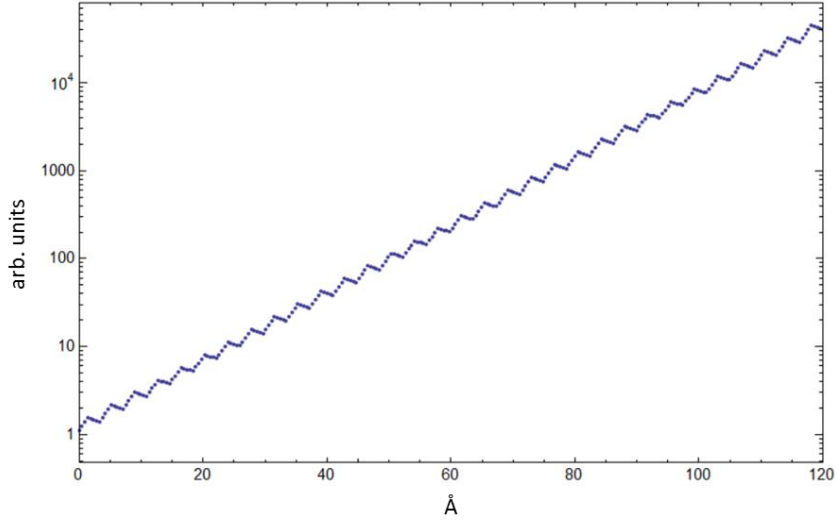

**Figure A1. Correlation function $\langle \chi_0^{1s} | \psi(z) \rangle$ within the quantum diffusion propagation.**

The QD correlation function was obtained using Sr *1s* $\chi_0^{1s}$ as a reference state and a wide Gaussian as an input $\psi(0)$. The resulting graph is purely real-valued and represents a monoexponential increase of all $c_n^\kappa$ associated with the Sr $\phi_{1s}$. The tooth-saw modulation arises due to piecewise constant potentials and show mainly the highest amplitude containing superposition of $\chi_{-1}^{1s}$ and $\chi_1^{1s}$ modes, which form a standing sinusoidal wave. To obtain $\chi_0^{1s}(r)$ wavefunctions with the least contribution from higher harmonics, the trick is to stop the QD evolution process at a value *z* which corresponds to the minima, i.e. where the sinusoid of the superimposed modulations is ~0. The quasi-energy $\varepsilon_\kappa$ can be extracted by fitting the QD correlation function, or approximately computed using Eq. (2.9).

## A2. Fitting of bound state-correlation functions

The real and imaginary parts of the state-correlation functions of Sr and Ti-O *1s*, obtained from simulations, are shown in Figure A2. Tooth-saw appearance arises from the piecewise constant Hamiltonians used in the multisliced propagation. The correlation functions were fitted in Fig. A2 b) according to:

$$\mathfrak{C}_\kappa(z) = \sum_n a_n^\kappa \, Cos[n g_z z + \theta_n] \, e^{-i\varepsilon_\kappa \sigma z} \, e^{-\sigma \Gamma_\kappa z} \, e^{i\theta_\kappa} \qquad (A2.1)$$

The fitting constants sought after are phases $\theta_n$, $\theta_\kappa$. Real and imaginary parts of the eigenvalues, $\varepsilon_\kappa = -112.7$ and $\sigma\Gamma_\kappa = 0.00074$, are extracted before hand via the spectral technique and fitting of excitation coefficient decays, respectively. The effective weigthing coefficients $a_n^\kappa$ are real-valued here, obtained from the spectra and taken as constant from the onset $z=0$. Assuming that $a_n^\kappa$ are constant rather than being populated within the transition interval $\bar{z}$ causes some issues with normalization. To match Fig. A2 b) with A2 a) we introduce an *ad-hoc* overall phase-factor $\theta_\kappa = \pi/20$ to Sr *1s* state in Fig. A2 b). It is then found that the sinusoid of *n=1* modes is out of phase from *n=0* by $\theta_1 = -\pi/2$, while *n=2* sinusoid shows $\theta_2 = \pi$ difference. This agrees with an inspection of *2p* wavefunctions obtained via the spectral filtering method (Fig. 3 c,f).

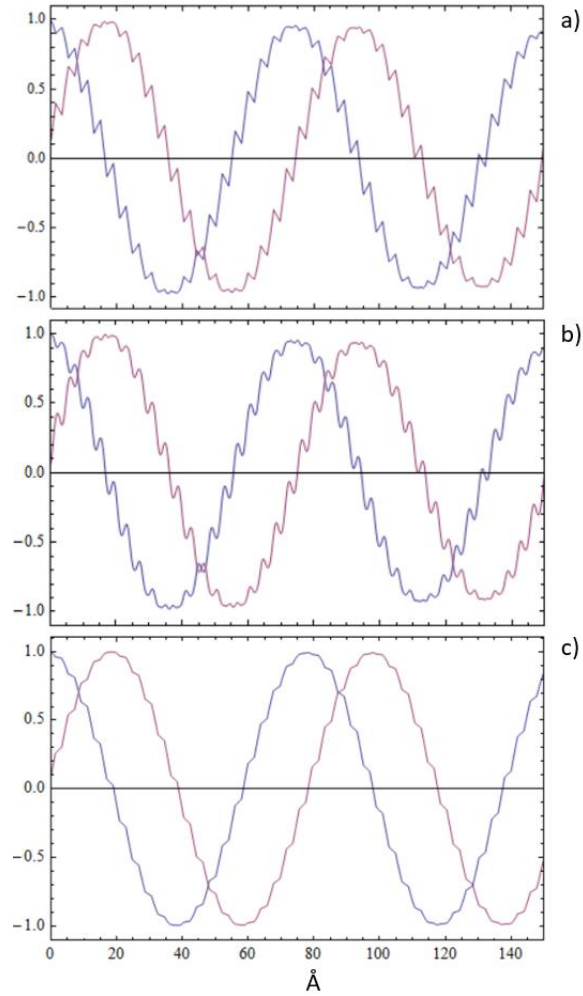

**Figure A2. a) Real (blue) and imaginary (red) parts of the Sr *1s* state-correlation function $\mathfrak{C}_\kappa(z) = \langle \chi_0^{1s} | \psi(z) \rangle$. b) Analytical fit using Eq. (A2.1) to Fig. a). c) Numerical state-correlation function of Ti-O *1s*-state showing smaller expansion coefficients of the higher modes, as compared to Sr *1s*.**

## A3. Fundamental Bloch wave eigensystem

For completeness we quote here the Bloch wave eigensystem describing dynamical high-energy electron propagation in crystals. Starting from the fundamental equations:

$$\left[K^2 - (\boldsymbol{k}^j + \boldsymbol{g})^2\right]C_g^j + \sum_{h \neq g} U_{gh} C_h^j = 0, \tag{A3.1}$$

the electron wavevector inside the crystal is customarily written as: $\boldsymbol{k}^j = \boldsymbol{K} + \varepsilon^j \boldsymbol{n}$, where $\boldsymbol{n}$ is unit vector normal to the surface, and $\boldsymbol{K}$ is the incident wavevector corrected for the inner-potential. The first-order eigensystem ignoring back-scattering is then obtained as:

$$2K_z S_g C_g^j + \sum_{h \neq g} U_{gh} C_h^j = 2K_z(1 + g_z/K_z)\varepsilon^j C_g^j \tag{A3.2}$$

Due to the $g_z/K_z$ term the matrix eigensystem in Eq. (A3.2) is non-Hermitian. It is a safe approximation to ignore this term in situations where bound states are excited very weakly. Otherwise, a non-orthogonal transformation of the eigenvectors $C_g^j$ can be introduced that renders the eigensystem symmetric, defining bi-orthogonal left-hand and right-hand eigenvectors (see for example [7]).

**References:**

[1] Dahmen, U., Erni, R., Radmilovic, V., Ksielowski, C., Rossell, M.-D., Denes, P., "Background, status and future of the transmission electron aberration-corrected microscope project", Philos. Trans. Royal Soc. Lond. A Math. Phys. Eng. Sci. **367**(1903), 3795–3808 (2009)

[2] Juan C. Idrobo and Stephen J. Pennycook, "Vortex beams for atomic resolution dichroism", Journal of Electron Microscopy 60(5): 295–300 (2011)

[3] Shibata, N., Findlay, S.D., Kohno, Y., Sawada, H., Kondo, Y., Ikuhara, Y. "Differential phase-contrast microscopy at atomic resolution", Nat. Phys. **8**(8), 611–615 (2012).

[4] Joanne Etheridge, Sorin Lazar, Christian Dwyer, and Gianluigi A. Botton, "Imaging High-Energy Electrons Propagating in a Crystal", PRL 106, 160802 (2011).

[5] Mark W. Tate et al., "High Dynamic Range Pixel Array Detector for Scanning Transmission Electron Microscopy", Microscopy and Microanalysis **22**(1), 237-249, (2016).

[6] Axel Lubk, Laura Clark, Giulio Guzzinati, Jo Verbeeck, "Topological analysis of paraxially scattered electron vortex beams", PHYSICAL REVIEW A 87, 033834 (2013)

[7] L.M. Peng, S.L. Dudarev, M.J. Whelan., "High-energy electron diffraction and microscopy", Oxford University Press, (2011).

[8] J. M. Cowley, A. F. Moodie, "The scattering of electrons by atoms and crystals. I. A new theoretical approach", Acta Cryst. (1957). 10, 609-619.

[9] Findlay, S., Allen, L., Oxley, M., Rossouw, C.: Lattice-resolution contrast from a focused coherent electron probe. Part II. Ultramicroscopy **96**(1), 65–81 (2003).

[10] E.J. Kirkland, "Advanced computing in electron microscopy", Plenum Press, New York (1998).

[11] Takao Morimura, "STEM image simulation by Bloch-wave method with layer-by-layer representation", Journal of Electron Microscopy 59, (Supplement): S23–S28 (2010)

[12] L.J. Allen, S. D. Findlay, M. P. Oxley, C. Witte, N. J. Zaluzec, "Channeling effects in high-angular-resolution electron spectroscopy", Phys Rev B 73, 094104 (2006)

[13] Colin Ophus, "A fast image simulation algorithm for scanning transmission electron microscopy", Adv Struct Chem Imag (2017), 3:13.

[14] P.D. Nellist and S.J. Pennycook, "The Principles and Interpretation of Annular Dark-Field Z-Contrast Imaging, Advances in imaging and electron physics, Vol 113.

[15] D. E. Jesson and S. J. Pennycook, "Incoherent Imaging of Crystals Using Thermally Scattered Electrons", Proceedings: Mathematical and Physical Sciences, Vol. 449, No. 1936 (May 9, 1995), 273-293

[16] S.D. Findlay, N.Shibata, H.Sawada, E.Okunishi, Y.Kondo, Y.Ikuhara, "Dynamics of annular bright field imaging in scanning transmission electron microscopy", Ultramicroscopy 110, (2010) 903–923

[17] G.R. Anstis, D.Q. Caib, D.J.H. Cockayne, "Limitations on the s-state approach to the interpretation of sub-angstrom resolution electron microscope images and microanalysis", Ultramicroscopy 94, (2003) 309–327

[18] P. Geuens, D. Van Dyck, "The S-state model: a work horse for HRTEM", Ultramicroscopy 93, (2002) 179–198

[19] J.M. Cowley, "Electron Microscopy of Crystals with Time-Dependent Perturbations", Acta Cryst. (1988). A44, 847-853

[20] C. Fanidis, D. Van Dyck and J. Van Landuyt, "Inelastic scattering of high-energy electrons in a crystal in thermal equilibrium with the environment II. Solution of the equations and applications to concrete cases", Ultramicroscopy 48, (1993) 133-164

[21] J.H. Chen, D. Van Dyck, M. Op de Beeckl, J. Van Landuyt, "Computational comparisons between the conventional multislice method and the third-order multislice method for calculating high-energy electron diffraction and imaging", Ultramicroscopy 69, (1997) 219-240

[22] Bing K. Jap, R.M. Glaeser, "The Scattering of High-Energy Electrons. I. Feynman Path-Integral Formulation", Acta Cryst. (1978). A34, 94-102

[23] J.J. Sakurai, "Modern Quantum Mechanics", Revised Edition. Pearson Education, 1994.

[24] H. Sambe, "Steady States and Quasienergies of a Quantum-Mechanical System in an Oscillating Field", Phys. Rev. A 7, 6, (1973).

[25] U. Peskin, N. Moiseyev, "Time-independent scattering theory for time-periodic Hamiltonians: Formulation and complex-scaling calculations of above-threshold-ionization spectra", Phys Rev A, 49, 5, (1994).

[26] P. Brieta, C. Fernandez, "Exponential decay and resonances in a driven system", Journal of Mathematical Analysis and Applications 396, (2) 513-522, (2012).

[27] A. Galtbayar, A. Jensen, K. Yajima, "Local time-decay of solutions to Schrodinger equations with time-periodic potentials", Journal of Statistical Physics 116, 1/4, (2004).

[28] T. Kovar, P. Martin, "Scattering with a periodically kicked interaction and cyclic states", J. Phys. A. Math. Gen. 31, 385-396, (1998).

[29] J. Augustin, A. Schafer, W. Greiner. Physical Review *A* **51,** 1367-1373 (1995).

[30] J.U. Andersen *et al*, "Axial channelling radiation from MeV electrons", Nuclear Instruments and Methods 194, 209-224, (1982).

[31] N.W. Ashcroft, N.D. Mermin, "Solid State Physics", Cengage Learning, Inc, (1976).

[32] P.A. Doyle, P.S. Turner, "Relativistic Hartree-Fock X-ray and electron scattering factors", Acta Cryst. A**24**, 390-397 (1968).

[33] I. Lobato, D.Van Dyck, "A complete comparison of simulated electron diffraction patterns using different parameterizations of the electron scattering factors", Ultramicroscopy 155, (2015) 11–19.

[34] Jones, P., Rackham, G., Steeds, J., "Higher order laue zone effects in electron diffraction and their use in lattice parameter determination", Proc. Royal Soc. Lond. A Math. Phys. Eng. Sci. **354**, 197–222 (1977)

[35] L.M. Peng *et al*, "Debye-Waller factors and absorptive scattering factors of elemental crystals", Acta Cryst. A52, 456-470, (1996).

[36] M.D. Feit, J.A. Fleck, A. Steiger, "Solution of the Schrodinger equation by the spectral method", Journal of Computational Physics **47**, 412-433 (1982).